\documentstyle[12pt,epsf]{article}
\begin{document}
\date{}
\begin{flushright}
{\bf CU-TP-1031}
\end{flushright}
\vskip20pt
\begin{center}
{\large\bf A Simple Derivation of the JIMWLK Equation}
\end{center}

\vskip 20pt
\begin{center}
{A.H. Mueller\footnote{This work is supported in part by the Department of Energy.}\\
Physics Department, Columbia University\\
New York, N.Y. 10027}
\end{center}

\noindent{\bf Abstract}
\vskip 20pt
A simple derivation of the Jalilian-Marian, Iancu, McLerran, Weigert, Leonidov and Kovner (JIMWLK)
equation for the evolution of small-x QCD wavefunctions is given.  The derivation makes use of the equivalence between the evolution of a (in
general complicated) small x wavefunction with that of the evolution of the (simple) dipole probing the wavefunction in a high energy
scattering.

\section{Introduction}

One expects the light-cone wavefunction of a very high energy hadron to exhibit saturation\cite{Gri}.  That is, the occupation number, in
transverse phase space, of quarks and antiquarks of momentum fraction x should be near 1 for transverse momentum $k_\perp< Q_s(x)$ while that of
gluons should be of size $1/\alpha.$  $Q_s,$ the saturation momentum, depends on the hadron in question and on the longitudinal momentum
fraction of the parton being considered.  $Q_s$ values of about $1  GeV,$ or perhaps a bit larger, may be available for study at HERA and at RHIC
and it is extremely pressing to find out if saturation is clearly visible at these accelerators.

Over the past 7 years or so there has been an ambitious program dedicated to finding appropriate equations for dealing with such high density
QCD wavefunctions\cite{L.Mc,J.Ja,Jal,Wei,Ian,McLe}.  This program has been quite successful and a renormalization group equation in the
form of a functional Fokker-Planck equation for the wavefunction of a high
energy hadron has been given by the authors of Refs.(4-6) (JIMWLK).  The most complete derivation is given in Ref.(6) where
the equation is written in terms of a covariant gauge potential, $\alpha,$ coming from light-cone gauge quanta in a high energy hadron.  The
mixture of light-cone gauge and covariant gauge comes about as a mathematical technique for simplifying an equation which appears to be
exceedingly complicated when expresssed purely in terms of light-cone gauge quantities.

It has been known for sometime that there is a dual description where the unitarity limit for scattering of a small dipole on a high energy
hadron is equivalent to saturation\cite{a.h.mu,muell}.  In this dual picture one views dipole-hadron scattering in a frame, the dipole frame,
where the dipole is left-moving and the hadron is right-moving.  The dipole rapidity, $y,$ is kept small enough, $\alpha_sy << 1,$ so that one
need not consider radiative corrections to the dipole wavefunction.  In a high energy scattering the rapidity of the hadron is very large and
the hadrons wavefunction involves many gluons.  If, at a given impact parameter, the S-matrix for dipole-hadron scattering is significantly
less than 1 then this strong scattering is equivalent to the statment that the occupation number of quarks and antiquarks, at a transverse
momentum corrresponding to the inverse dipole size, is of the order of 1.

In this paper we use the dipole frame to derive the JIMWLK equation in a rather simple manner.  As one increases the
energy of a high energy dipole-hadron scattering by a small amount, this increase can either be put into the hadron or into the dipole.  If
the increase is put into the  hadron then the wavefunction of the hadron changes (evolves).  If the increase is put into the dipole the dipole
can emit a gluon in an easily calculable manner.  By equating these two pictures one arrives at the JIMWLK equation.  In a sense our physical
picture is a simple way of viewing the set of mathematical transformations carried out in Ref.(6) in their derivation of this equation. 

\section{The JIMWLK Equation}

 Let

\begin{equation}
V^\dag({\underline x}) = P  exp\{ig\int_{-\infty}^\infty dx_-A_+({\underline x},x_-)\}
\end{equation}

\noindent be a path-ordered line integral taken, say, at $x_+ = 0$ where

\begin{equation}
A_+(x) = \sum_{a=1}^{N_c^2-1} T^aA_+^a(x)
\end{equation}

\noindent with $T^a$ the generators of color $SU(N_c).$  The symbol  $P$  in (1) indicates that matrices $A_+({\underline x}, x_-)$ having larger values of
$x_ -$ are put to the left of those having smaller values of $x_-.$  In a moment we shall specialize our discussion to Coulomb gauge, but for now any
gauge for which
$A_+$ is not identically zero and for which
$A_i({\b  x}, x_-)_{\longrightarrow\atop_{x_-\rightarrow\pm\infty}}0$ is suitable.  For simplicity of presentation we take the $T^a$ to be in the
fundamental representation of $SU(N_c).$

Let $O(A_+)$ be a gauge invariant functional formed out of $V^\prime s$ and $V^{\dag\prime}s$ at various values of ${\underline x}.$  For example,

\begin{equation}
O_2({\underline x}_1,{\underline x}_2) = tr V^\dag({\underline x}_1) V({\underline x}_2)
\end{equation}

\noindent and

\begin{equation}
O_4({\underline x}_1, {\underline x}_2, {\underline x}_3, {\underline x}_4) = tr V^\dag({\underline x}_1) V({\underline x}_2) V^\dag({\underline x}_3) V({\underline x}_4)
\end{equation} 

\noindent constitute such functionals so long as $A_i$ vanishes at $x_-=\pm\infty.$

Suppose $\vert p >$ is the state of a hadron having momentum

\begin{equation}
(p_0, {\underline p}, p_z) = M(cosh\ Y, 0, sinh\  Y)
\end{equation}

\noindent where the rapidity  $Y$ is large and positive.  Then one may write\cite{L.Mc}

\begin{equation}
<O>_Y = <p\vert O\vert p> = \int {\Large\bf\it D}[\alpha({\underline x}, {x}_-)] O(\alpha) W_Y[\alpha]
\end{equation}

\noindent where the weight function $W_Y$ is given by

\begin{equation}
W_Y[\alpha]=\int{\Large\bf\it D} [A_\mu] \delta(A_+-\alpha) \delta(F(A))\Delta_F[A]e^{iS[A]}.
\end{equation} 

\noindent The symbol ${\Large\bf\it D}$ indicates a functional integral, $F$ is a gauge fixing, and $\Delta_F$ is the corresponding
Fadeev-Popov determinant times an operator which projects out the state $\vert p >.$  We suppose $W_Y,$ the distribution of values of the potential $A_+$ in the hadron
wavefunction, is normalized to

\begin{equation}
\int {\Large\bf\it D}[\alpha]W_Y[\alpha] = 1
\end{equation}

To interpret (6) more completely we now specify our gauge choices as Coulomb gauge.  The advantage of Coulomb gauge\cite{T.Jar} is that it gives a nice
separation between left-moving quanta, right-moving quanta and gluons exchanged between left and right-movers.  Quanta having $k_+/k_->>0$ appear as they
would in $A_+=0$ gauge.  Quanta having $k_-/k_+>>1$ appear as they would in $A_-=0$ gauge.  Exchanged gluons appear as in covariant gauge if $k_+, k_-
<< k_\perp.$  Thus $<O({\underline x}_1,{\underline x}_2)>_Y$ corresponds to the S-matrix for scattering of a left-moving dipole, of size ${\underline x}_1-{\underline x}_2$ on a
right-moving hadron.  $W_Y[\alpha]$ gives the field $A_+=\alpha$ coming from all the right-moving quanta in the high energy hadron.  The rapidity of the
scattering is $Y.$  This latter statement is, perhaps, not immediately clear since $O_2$ is defined in terms of left-movers exactly on the light cone
$x_+=0.$  However, in order that $O_2$ correctly represent a left-moving dipole we must assume that its rapidity, $y,$ obeys $\alpha_sy<<1.$  If
$\alpha_sy\geq 1$ the bare dipole will become dressed with additional left-moving gluons so that $O_2$ would not represent a physical dipole.

Now consider the $Y$-dependence of $<O>_Y.$  From (6) one can write

\begin{equation}
{d\over dY} <O>_Y = \int {\Large\bf\it D}[\alpha] O(\alpha){d\over dY} W_Y[\alpha].
\end{equation}

\noindent However, one can equally well imagine calculating ${d\over dY} <O>_Y$ by evaluating the change of the left-moving dipole, due to an additional
gluon emission, as one increases the rapidity of the left-moving system by an amount $dY.$  The change in the left-moving state is given by the gluon
emission and absorptions shown in Figs.1 and 2 where spectator quark and antiquark lines are not shown.  In Fig.1 we have assumed that a gluon line
connects a quark and an antiquark.  This is for the sake of definiteness.  We could have equally well assumed a connection between two quarks or
between two antiquarks.  In Fig.2 we suppose that the gluon is emitted and absorbed on the same quark (or antiquark) line.  The vertical lines in the
figures represent the ``time,'' $x_-=0,$ at which the left-moving system passes over the right-moving hadron and its gluon fields.  This view of the
$Y-$dependence of $<O>_Y$ in terms of a change of the rapidity of the (rather simple) left-movers is in the spirit of work previously done by
Balitsky\cite{I.Bal}, Kovchegov\cite{Yu.V}, and Weigert\cite{Wei}.

\begin{center}
\begin{figure}
\epsfbox[0 0 456 96]{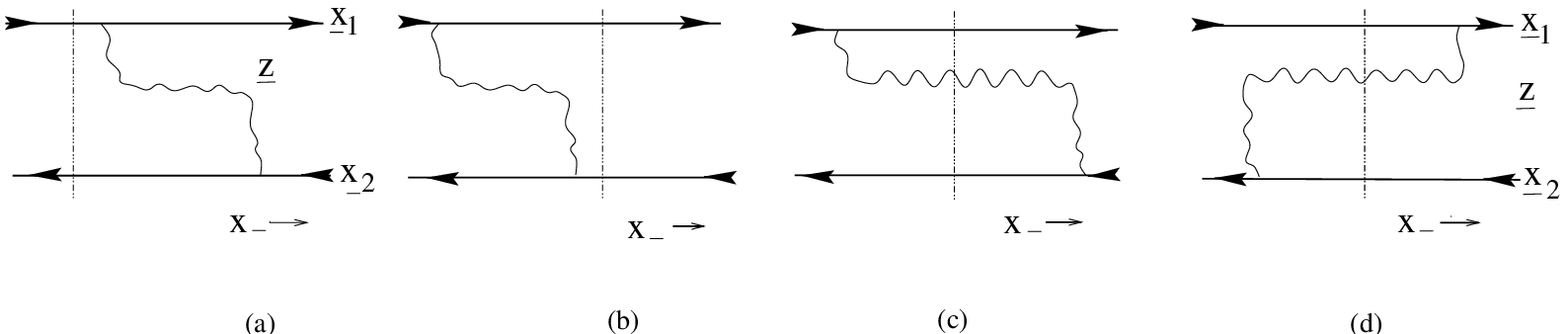}
\bigskip
\center{FIG.1}
\end{figure}
\end{center}

Let's examine the contribution of the graph of Fig.1a to ${d\over dY} <O>_Y$ in more detail.  In this contribution a gluon is emitted off a quark at
${\underline x}_1$ and absorbed by an antiquark at ${\underline x}_2.$  The emission and absorption happen after the left-moving system has passed the hadron so the
gluon does not interact with the hadron.  We do the calculation in $A_-=0$ light-cone gauge which for left-movers is equivalent to Coulomb gauge. 
Including the term where the gluon is first emitted off the antiquark and then absorbed by the quark one has the simple replacement\cite{373}.

\begin{equation}
V^\dag({\underline x}_1) \otimes V({\underline x}_2) \rightarrow {\alpha\over \pi^2} \int d^2z {({\underline x}_1-{\underline z})\cdot ({\underline x}_2-{\underline z})\over
({\underline x}_1-{\underline z})^2({\underline x}_2-{\underline z})^2} T^cV^\dag({\underline x_1})\otimes V({\underline x}_2)T^c.
\end{equation}

\noindent in going from $O$ to ${d\over dY}O.$  The notation $V^\dag\otimes V$ means that the color indices on $V$ and $V^\dag$ are fixed with no matrix
multiplication implied.  Of course $T^cV^\dag({\underline x}_1)$ does involve a matrix multipication of $T^c\  {\rm times}\  V^\dag({\underline x}_1)$ on the right-hand side of
(10).  The fact that $T^c$ comes to the left of $V^\dag{\underline x}_1)$ but to the right of $V({\underline x}_2)$ in (10) follows from the $x_--$ ordering given in (1)
and the fact that the contribution of Fig.1a has gluon emission and absorption at (large) positive values of $x_-,$ that is after the scattering has
taken place.

Now consider the contribution illustrated in Fig.1c.  This contribution involves the replacement

\begin{equation}
V^\dag({\underline x}_1) \otimes V({\underline x}_2) \rightarrow {-\alpha\over \pi^2} \int d^2z {({\underline x}_1-{\underline z})\cdot ({\underline x}_2-{\underline z})\over
({\underline x}_1-{\underline z})^2({\underline x}_2-{\underline z})^2} {\tilde V}_{cd}({\underline z}) V^\dag({\underline x}_1) T^c\otimes V({\underline x_2})T^d.
\end{equation}

\noindent The new elements which arise in (11) compared to (10) are the following: (i)  The overall sign is (-) because the gluon has been emitted
before the scattering and absorbed after the scattering.  (ii) The factor ${\tilde V}_{cd}({\underline z})$ gives the interaction of the gluon with the
right-moving hadron as an eikonal ``phase'' as in (1) but now in the adjoint representation. (iii)  Now the $T^c$ come to the right of $V^\dag({\underline
x}_1)$ because the gluon is emitted off the ${\underline x}_1-$ line before the interaction with the hadron.  For reason which should become clear shortly it is
useful to write

\begin{equation}
V^\dag({\underline x}_1) T^c = V^\dag({\underline x}_1)T^cV({\underline x}_1) V^\dag({\underline x}_1) = {\tilde V}_{ca}({\underline x}_1) T^aV^\dag({\underline x}_1)
\end{equation}

\noindent in order to bring the $T$ to the left of $V^\dag({\underline x}_1).$  Thus one can write, replacing (11),

\begin{displaymath}
V^\dag({\underline x}_1)\otimes V({\underline x}_2)\rightarrow {-\alpha\over \pi^2}\int d^2{\underline z}{({\underline x}_1-{\underline z}) \cdot ({\underline x}_2-{\underline
z})\over ({\underline x}_1-{\underline z})^2 ({\underline x}_2-{\underline z})^2}\cdot
\end{displaymath}
\begin{equation}
\cdot [{\tilde V}^\dag({\underline x}_1) {\tilde V}({\underline z})]_{ab}
T^aV^\dag({\underline x}_1) \otimes V({\underline x}_2)T^b.
\end{equation}
\vskip 5pt
\noindent There is a simple mnemonic for seeing the structure of the $V^\prime$s and ${\tilde V^\prime}$s in (13).  Instead of the graph in Fig.1c we
could use that shown in Fig.3 where an adjoint line integral starts at large positive values of $x_-$ follows the quark line at ${\underline x}_1$ as $x_-$
decreases to large negative values and then proceeds from large negative to large positive values of $x_-$ at the point ${\underline z}.$  The interaction of
the hadron with the quark and gluon, both at ${\underline x}_1,$ in Fig.3 is the same as with the quark line in Fig.1c.

It is now straightforward to add in the contributions of the graphs shown in Fig.1b and 1d to get the total contribution of the graphs of Fig.1 as

\begin{displaymath}
V^\dag({\underline x}_1) \otimes V({\underline x}_2) \rightarrow {\alpha\over \pi^2} \int d^2z{({\underline x}_1-{\underline z})({\underline x}_2-{\underline z})\over ({\underline
x}_1-{\underline z})^2({\underline x}_2-{\underline z})^2}\cdot
\end{displaymath}

\begin{equation}
\cdot\{1+{\tilde V}^\dag({\underline x}_1){\tilde V}({\underline x}_2)-{\tilde V}^\dag({\underline x}_1){\tilde V}({\b z})-{\tilde V}^\dag({\underline z}){\tilde V}({\underline
x}_2)\}_{ab}
\cdot T^aV^\dag({\underline x}_1)\otimes V({\underline x}_2) T^b,
\end{equation}

\noindent when $O \rightarrow {d\over dY}O.$  This can be written as

\begin{equation}
V^\dag({\underline x}_1) \otimes V({\underline x}_2) \rightarrow {\alpha\over 2} \int d^2 x d^2y \eta_{{\underline x}{\underline y}}^{ab}\  {\delta^2\over
\delta\alpha^a({\underline x}, x_-)
\delta\alpha^b({\underline y}, y_-)} V^\dag({\underline x}_1)\otimes V({\underline x}_2)
\end{equation}

\noindent with\cite{Ian}

\begin{displaymath}
g^2\eta_{{\underline x}{\underline y}}^{ab}= 4\int {d^2z\over (2\pi)^2} 
{({\underline x}-{\underline z})\cdot ({\underline y}-{\underline z})\over ({\underline x} - {\underline z})^2 ({\underline y}-{\underline z})^2}\cdot
\end{displaymath}

\begin{equation}
\cdot\{1 + {\tilde
V}^\dag({\underline x}){\tilde V}({\underline y)-{\tilde V}^\dag({\underline x}){\tilde V}({\underline z})-{\tilde V}^\dag(\underline z})V{\underline y})\}_{ab}
\end{equation}
\vskip 5pt
\noindent where it is understood that the two functional derivatives on the right-hand side of (15) do not both act on either $V$ or on $V^\dag.$

One might expect that the contributions of the graphs shown in Fig.2 would come from an equation like (15) but where the two functional derivatives act
on a single term, say $V^\dag({\underline x}_1),$ and where the other factor, $V({\underline x}_2)$ is absent.  This is almost the case.  The contributions illustrated in
Figs.2a and 2b do correspond to an equation like (15) taking only the first two terms in the brackets for the $\eta$ given in (16).  However,

\begin{center}
\begin{figure}
\epsfbox[0 0 396 64]{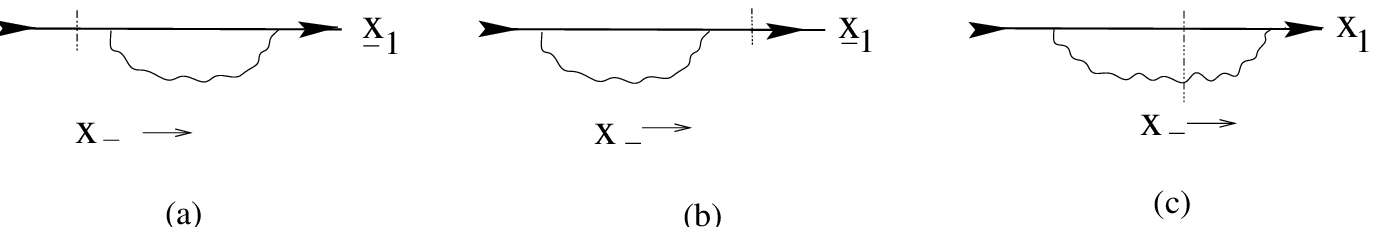}
\bigskip
\center{FIG.2}
\end{figure}
\end{center}

\begin{displaymath}
{-1\over 8\pi^3}\int d^2xd^2y  d^2z {({\underline x}-{\underline z})({\underline y}-{\underline z})\over ({\underline x}-{\underline z})^2(y-{\underline z})^2}\{-{\tilde
V}^\dag({\underline x}){\tilde V}({\underline z})-{\tilde V}^\dag({\underline z}){\tilde V}({\underline y})\}_{ab}
{\delta^2\over \delta\alpha^a({\underline x},x_-)\delta\alpha^b({\underline y},y_-)}\cdot
\end{displaymath}
\begin{equation} 
\cdot V^\dag({\underline x}_1) = {\alpha\over 2\pi^2} \int {d^2z\over ({\underline x}_1-{\underline z})^2}
\{{\tilde V}^\dag({\underline x}_1)
{\tilde V}^\dag({\underline z}) + {\tilde V}^\dag({\underline z}){\tilde V}({\underline
x}_1)\}_{ab}T^aT^bV^\dag({\underline x}_1)
\end{equation}

\vskip 5pt
\noindent while the contribution of the term shown in Fig.2c is 
\vskip 5pt
\begin{equation}
{\alpha\over \pi^2} \int {d^2z\over ({\underline x}_1-{\underline z})^2} T^bV^\dag({\underline x}_1) T^a {\tilde V}_{ab}({\underline z}) = {\alpha\over \pi^2} \int {d^2z\over
({\underline x}_1-{\underline z})^2} T^bT^a
\{{\tilde V}^\dag ({\underline x}_1){\tilde V}({\underline z})\}_{ab}{\tilde V}^\dag({\underline x}_1). 
\end{equation}

\begin{center}
\begin{figure}
\epsfbox[0 0 200 103]{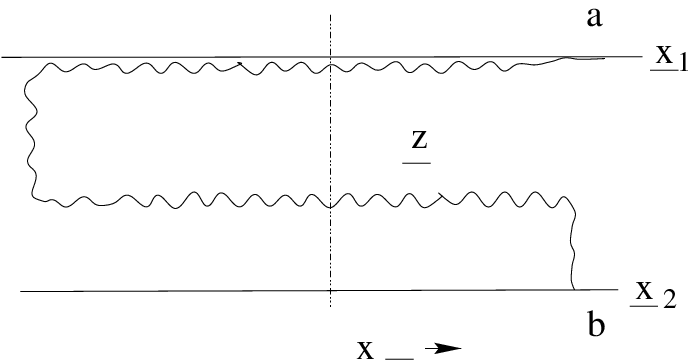}
\bigskip
\center{FIG.3}
\end{figure}
\end{center}

\noindent The difference between the expressions on the right-hand sides of (18) and (17) is

$$
{\alpha\over 2\pi^2}\int {d^2z\over ({\underline x}_1-{\underline z})^2} \{{\tilde V}^\dag({\underline x}_1){\tilde V}({\underline z})-{\tilde V}^\dag({\underline z}){\tilde
V}({\underline x}_1)\}_{ab}T^bT^aV^\dag({\b x}_1)
$$
\begin{equation}
=\ -\ {\alpha\over 2\pi^2} \int {d^2z\over ({\underline x}_1-{\underline z})^2} tr[T^c{\tilde V}^\dag({\underline x}_1){\tilde V}({\underline z})]T^cV^\dag({\underline x}_1)
\end{equation}
\noindent Thus taking (15) and (19) we can write

\begin{equation}
{dO\over dY} = {\alpha_s\over 2} \int d^2 xd^2 y \eta_{{\underline x}{\underline y}}^{ab} {\delta^2\over\delta\alpha^a({\underline x}, x_-)\delta\alpha^b({\underline
y},y_-)} O(\alpha) + \alpha_s\int d^2x \nu_{_{\underline x}}^a {\delta\over
\delta\alpha^a({\underline x},{\rm x}_-)} O(\alpha)
\end{equation}

\noindent where\cite{Ian}

\begin{equation}
g\nu_{_{\underline x}}^a = 2i\int {d^2z\over ({\underline x}-{\underline z})^2} tr[T^a{\tilde V}^\dag({\underline x}_1){\tilde V}({\underline z})]
\end{equation}

\noindent and where, as usual, $x_-$ and $y_-$ are positive.

Finally, equating these two pictures of $Y-$evolution given by (9) and by (20) we can write

\begin{displaymath}
\int {\Large\bf\it D}[\alpha]O(\alpha) {dW_Y\over dY}[\alpha]= \int {\Large\bf\it D}[\alpha]O(\alpha)
\end{displaymath}
\begin{equation}
\alpha_s\{{1\over 2}\int d^2xd^2y{\delta^2\over \delta\alpha^a({\underline x}, x_-)\delta\alpha^b({\underline y}, y_-)}[W_Y\eta_{{\underline x}{\underline y}}^{ab}]-\int d^2x
{\delta\over \delta\alpha^a({\b x}, x_-)}[W_Y\nu_{\underline x}^a]\}.
\end{equation}
\vskip 5pt
\noindent where we have integrated by parts in going from (20) to the right-hand side of (22).

Eq.(22) can be cast into a form closer to that used in JIMWLK by defining, for ${\bar Y}-Y >>1,$

\begin{equation}
W_{{\bar Y},Y}[\alpha] = W_Y[{\tilde \alpha}]
\end{equation}

\noindent where

\begin{equation}
\alpha({\underline x}, x_-) = e^{{\bar Y}-Y}{\tilde \alpha}({\underline x}, e^{{\bar Y}-Y} x_-).
\end{equation}

\vskip 5pt
\noindent That is, $W_{{\bar Y},Y}[\alpha]$ corresponds to the same functional integral given in (7) but where the state $\vert p >$ now has
rapidity ${\b Y}$ and where only fields of longitudinal momentum $k_+/p_+ \geq e^{-Y}$ are included in the functional ingtegration.  Since the
right-moving fields are effectively in $A_+=0$ light-cone gauge in (7) the Lorentz boost indicated in  (23) and (24) is purely kinematic.  Since
$O(\alpha)$ in (22) constitutes a very general form of gauge invariant functional we presume, but do not know how to prove, that (22) - (24) imply

\begin{displaymath}
{d\over dY} W_{{\bar Y},Y}[\alpha]=\alpha_s\{{1\over 2} \int d^2xd^2y{\delta^2\over \delta\alpha^a({\underline x}, x_-) \delta\alpha^b({\underline y},y_-)}
[W_{{\bar Y},Y}\eta_{{\underline x}{\underline y}}^{ab}]
\end{displaymath}
\begin{equation}
-\int d^2x {\delta\over \delta\alpha^a({\underline x},x_-)}[W_{{\bar Y},Y}\nu_{\underline x}^a]\},
\end{equation}

\vskip 5pt
\noindent the form of the JIMWLK equation given in Refs. 6 and 7.

What are the values of $x_-$ and $y_-$ in (25), or equivalently, in (22)?  The derivation of (22) makes clear that all that is required is that $x_-$ and
$y_-$ be large enough so that

\begin{equation}
g \int_{x_-}^\infty \alpha({\underline x}, x_-^\prime) dx_-^\prime,\  g \int_{y_-}^\infty \alpha({\underline y},y_-^\prime) dy_-^\prime << 1 
\end{equation}

\noindent allowing (14) to be replaced by (15).  The exact values of $x_-$ and $y_-$ are not important although one should require\cite{Ian,McLe}

\begin{equation}
{1\over Q} e^{-({\bar Y}-Y)} << x_-, y_- << {1\over Q} e^{-({\bar Y}-Y-\Delta y)}
\end{equation}

\noindent with $\Delta y = 1/\alpha$  in order that there not be additional gluons emitted with rapidity between ${\b Y}- Y$ and the
rapidity of the gluon we are considering.  Q is a scale characterizing the typical transverse momentum of gluons at rapidity ${\b Y}-
Y$.  In terms of kinematics of the original frame used in (22) the requirement (27) is simply that $\alpha_s\ell n(x_-Q) <<1$ and $\alpha_s \ell
n(y_-Q)<<1$ similar to the condition imposed earlier in the rapidity of the left-moving dipole.

We have noted several times that the right-moving gluons determining\   W\   can be viewed as light-cone gauge, $A_+=0,$ quanta.  Operations, such as
indicated in (12), which bring generators always to the left of $V^\dag$ and to the right of $V$ make the further choice of an $A_+=0$ gauge where the
gauge fields vanish for $x_-<0,$ thus matching the choice of Refs.6 and 7.  Equally well one could have chosen to move all generators, $T^a,$ to the
right of $V^\dag$ and to the left of $V$ which would correspond to the light-cone gauge where potentials vanish, for high momentum particles, for
$x_->0$ as was done previously in Refs.14-16 in a different context.

\vskip 10pt
\noindent{\bf Acknowledgements}

I am grateful to Larry McLerran, Heribert Weigert and especially to Edmond Iancu for extremely informative and stimulating discussion on their work.


\begin{thebibliography}{99}
\bibitem{Gri}	L.V. Gribov, E.M. Levin and M.G. Ryskin, Phys.Rep.100 (1983)1.
\bibitem{L.Mc}L.Mc Lerran and R. Venugopolan, Phys. Rev.D49 (1994) 2233; Phys. Rev.D49 (1994) 3352; Phys. Rev.D50 (1994) 2225.	
\bibitem{J.Ja}	J.Jalilian-Marian, A. Kovner, L. McLerran and H. Weigert, Phys.Rev.D55 (1997) 5414.
\bibitem{Jal}J.Jalilian-Marian,A. Kovner, A. Leonidov and H. Weigert, Nucl. Phys. B504 (1997) 415; Phys. Rev.D59 (1999) 014014.
\bibitem{Wei}	H. Weigert, hep-ph/0004044.
\bibitem{Ian}	E.Iancu, A. Leonidov and L. McLerran, Nucl.Phys.A692 (2001) 583; Phys.Lett.B510 (2001) 133; hep-ph/0109115.
\bibitem{McLe} E. Iancu and L. McLerran, Phys.Lett.B510 (2001) 145. 
\bibitem{a.h.mu}	A.H. Mueller, and G.P. Salam, Nucl.Phys.B475 (1996) 293.
\bibitem{muell} A.H. Mueller, Nucl.Phys.B558 (1999) 285.
\bibitem{T.Jar} T. Jaroszewicz, Acta Physica Polonica B11 (1980) 965.
\bibitem{I.Bal} I. Baltisky, Nucl.Phys. B463 (1996) 99.
\bibitem{Yu.V} Yu.V. Kovchegov, Phys.Rev.D 60 (1999) 034008; Phys.Rev.D 61 (2000) 074018.
\bibitem{373} A.H. Mueller, Nucl.Phys.B4151 (1994) 373.
\bibitem{5445} Yu.V. Kovchegov, Phus.Rev.D 55 (1997) 5445.
\bibitem{Qiu} A.H. Mueller, and J. Qiu, Nucl.Phys. B268 (1986) 427.
\bibitem{451} Yu.V. Kovchegov and A.H. Mueller, Nucl.Phys.B529 (1998) 451. 
\end{thebibliography}
\end{document}